\documentclass[11pt,twoside]{article}
\usepackage{asp2010}

\begin{document}
\label{talk:ribeiro}

\title{On the Evolution of the Late-time {\it Hubble Space Telescope} Imaging of the Outburst of the Recurrent Nova RS Ophiuchi (2006)}
\author{Val\'erio A. R. M. Ribeiro,$^1$ Michael F. Bode,$^2$ and Robert E. Williams$^3$
\affil{$^1$Astrophysics, Cosmology and Gravity Centre, Department of Astronomy, University of Cape Town, Private Bag X3, Rondebosch, 7701, South Africa; vribeiro@ast.uct.ac.za}
\affil{$^2$Astrophysics Research Institute, Liverpool John Moores University, Egerton Wharf, Birkenhead, CH41 1LD, UK}
\affil{$^3$Space Telescope Science Institute, 3700 San Martin Drive, Baltimore, MD 21218, USA}}

%\aindex{Ribeiro,~V.~A.~R.~M.}
%\aindex{Bode,~M.~F.}
%\aindex{Williams,~R.~E.}

\begin{abstract}
We modelled the late-time {\it Hubble Space Telescope} imaging of RS Ophiuchi with models from Ribeiro et al. (2009), which at the time due to the unknown availability of simultaneous ground-based spectroscopy left some open questions as to the evolution of the expanding nebular from the early to the late time observations. Initial emission line identifications suggest that no forbidden lines are present in the spectra and that the emission lines arising in the region of the WFPC2 F502N images are due to N~{\sc ii} and He~{\sc i} $+$ Fe~{\sc ii}. The best model fit to the spectrum is one where the outer faster moving material expands linearly with time while the inner over-density material either suffered some deceleration or did not change in physical size. The origin of this inner over-density requires further exploration.
\end{abstract}

\section{Introduction}
RS~Oph is a member of the small group of objects known as symbiotic recurrent novae, with seven confirmed recorded outbursts. The system comprises a white dwarf, probably close to the Chandrasekhar limit (e.g. \citealt{DK94}) and a red-giant secondary of spectral type estimated around M0/2 III (e.g. \citealt{AM99}). The high mass white dwarf and high accretion rate is generally attributed to cause the short recurrence observed in these kinds of systems \citep{YPS05}. RS Oph's latest outburst, first detected on 2006 February 12.83 (\citealt{NHK06}), reached optical peak on 2006 February 12.94 \citep[][taken as $t=0$]{HBH10} and was observed from the hard X-ray to the radio (for recent reviews see \citealt{B10,R11}, and references therein).

The system was first resolved as a partial ring of non-thermal radio emission using Very Long Baseline Interferometry on day 13.8 after outburst. As the system evolved, a bipolar structure emerged. The asymmetry was suggested to be due to absorption in the overlying red-giant wind \citep{OBP06,OBB08}. Very Long Baseline Array imaging between days 34 $-$ 51 after outburst showed a central thermally dominated source linked to what appeared to be a collimated non-thermal outflow \citep*{SRM08}. \citet{TDP89} also interpreted their observations from day 77 after the 1985 outburst as a central thermal source and expanding non-thermal lobes.

\citet{RBD09} used combined {\it Hubble Space Telescope} ({\it HST}) imaging (using the ACS/HRC F502N filter) and ground-based spectroscopic observations on day 155 after outburst to compare with a model of RS Oph using a morpho-kinematical code to retrieve the underlying 3D structure. They modelled the ejecta as having a bipolar morphology composed of a high velocity outer dumbbell and a low velocity innermost hour-glass higher density structure. The innermost hour-glass structure replicated well the observed low velocity line width. \citet{RBD09} were also able to show that the asymmetry on the observed image was due to the finite width and offset of the central wavelength of the F502N filter. They derived the inclination of the system as 39$^{+1}_{-9}$ degrees and maximum expansion velocity of 5000$^{+1500}_{-100}$~km~s$^{-1}$. One key question that was left open was what happened to the morphology of the system as it evolved from day 155 to later epochs after outburst. Was it a linear expansion of both components or was there some other behaviour? This short contribution aims to shed light on this question, following archival data search, by modelling the additional {\it HST} imaging from day 449 and ground based spectroscopy (around the same date).

\section{Methods}
RS Oph was imaged with {\it HST} under the Director's Discretionary programs GO/DD-11075 on 2007 May 7 (day 449 after outburst) with the Planetary Camera CCD of the Wide Field Planetary Camera 2 (WFPC2), with a scale of 0.046$^{\prime\prime}$ pixel$^{-1}$. The {\it HST} data reduction steps have already been described in \cite{RBD09} and will not be detailed here.

An archival search of the Isaac Newton Group database\footnote{http://casu.ast.cam.ac.uk/casuadc/archives/ingarch} proved fruitful in finding a spectroscopic observation of RS Oph on 2007 April 03 (day 415 after outburst) with the 4.2 m William Herschel Telescope (WHT). The observation used the Intermediate Dispersion Spectrograph and Imaging System (ISIS), using grating R300B with a dispersion of 0.86 \AA/pix and a resolution $\sim$ 1000. The spectra were reduced using standard Image Reduction and Analysis Facility\footnote{IRAF is distributed by the National Optical Astronomical Observatories, which is operated by the Associated Universities for Research in Astronomy, Inc., under contract to the National Science Foundation.} procedures. However, no flux calibration was attempted as at this stage we are only interested in the line profile, but we used a standard star in order to remove the instrumental effect on the continuum.

\section{Results}
\begin{figure}[t]
\begin{center}
\includegraphics[width=\textwidth]{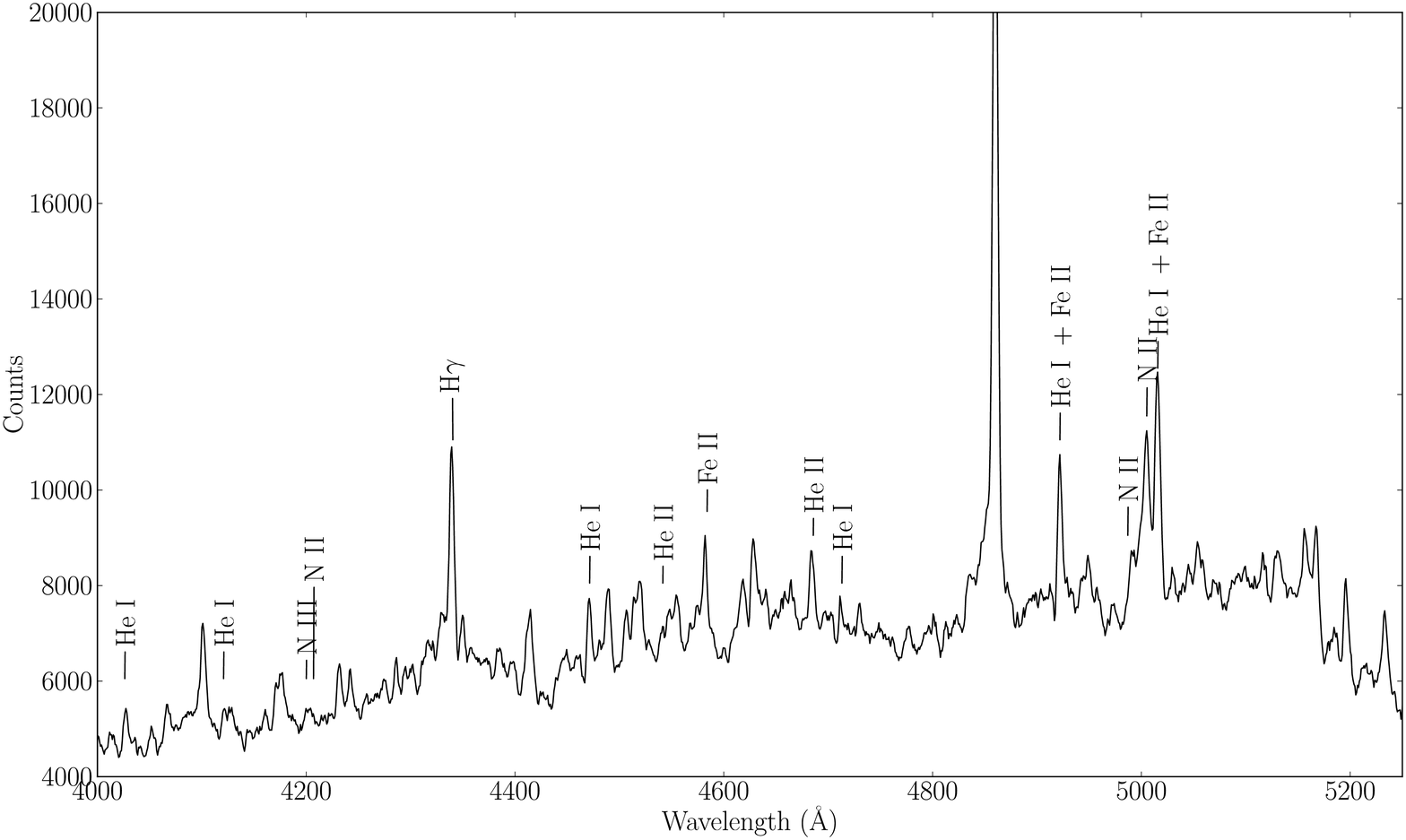}
\caption{Identification of emission lines in the WHT ISIS spectrum of RS Oph on day 415 after outburst. The data have not been flux calibrated. Clearly evident are the hydrogen Balmer lines, along with lines of nitrogen, helium and iron.}
\label{fig1}
\end{center}
\end{figure}

In Figure~\ref{fig1}, we show part of the optical spectrum (4000 -- 5250 \AA) with a few of the emission lines identified and what is clearly evident is that the spectrum is dominated by permitted lines of the hydrogen Balmer series, helium, nitrogen and iron. No forbidden lines are present. We should note that towards the red end of this spectrum the sensitivity is deminished considerably.

\begin{figure}[t!]
\begin{center}
\plottwo{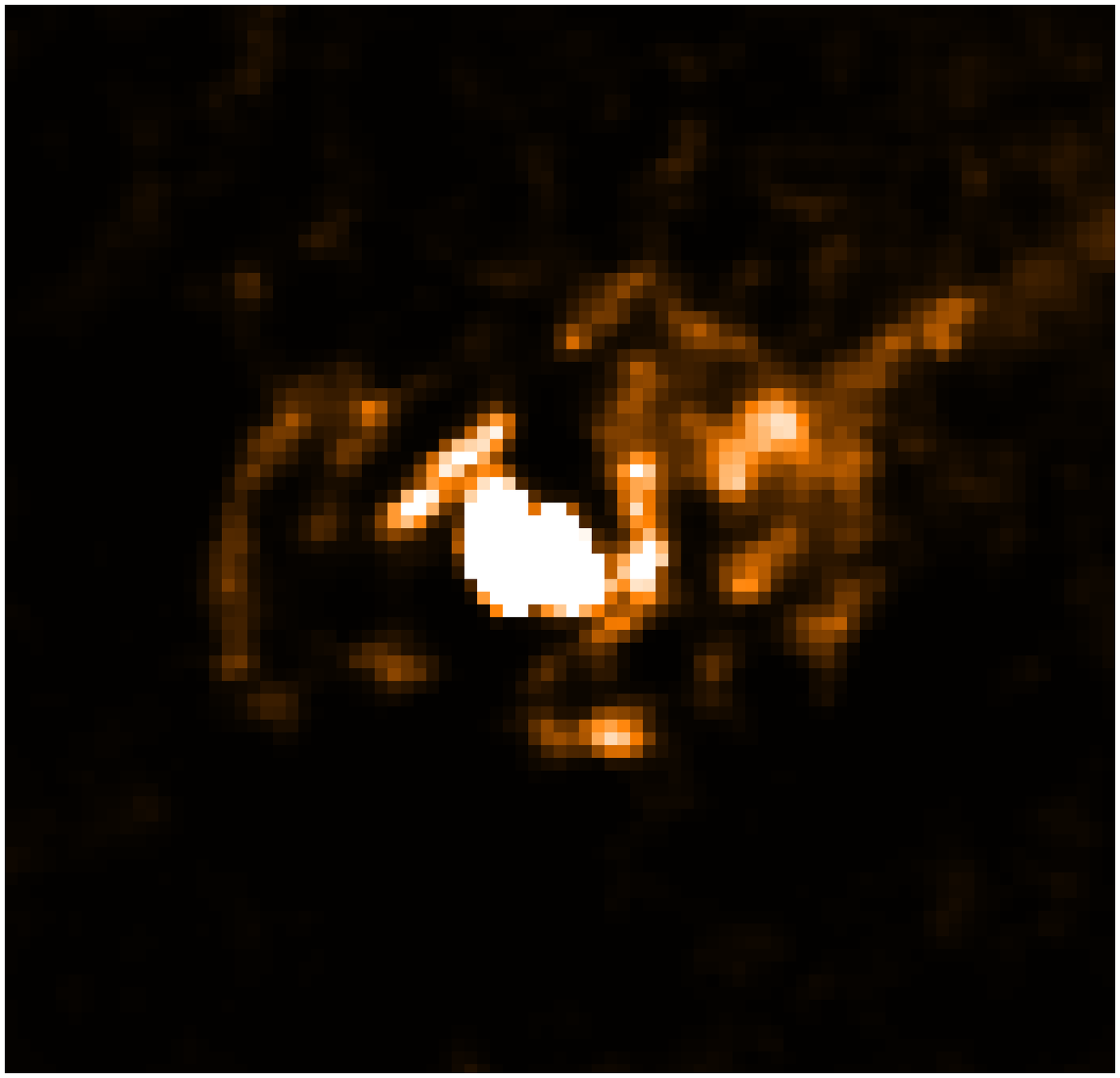}{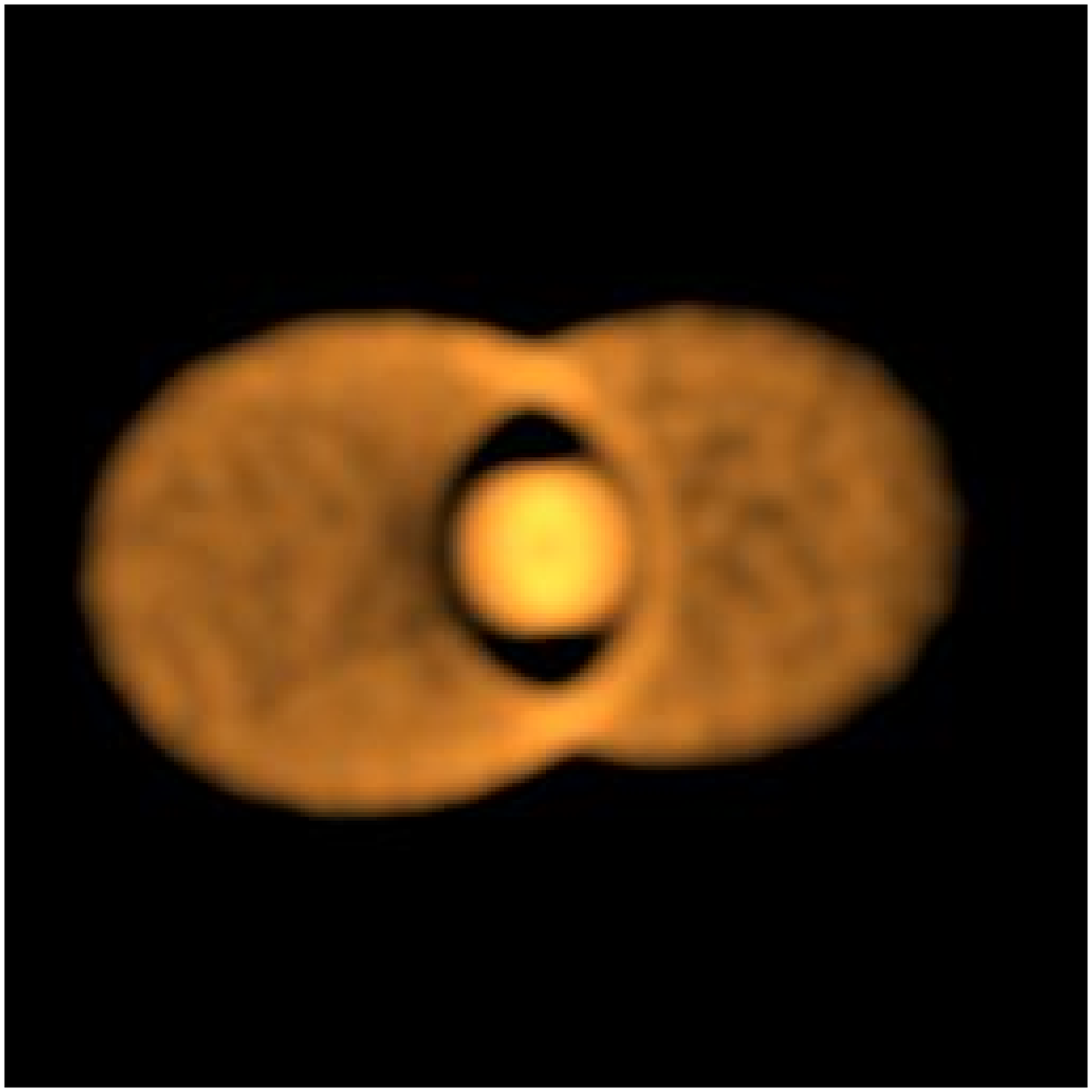}
\plotone{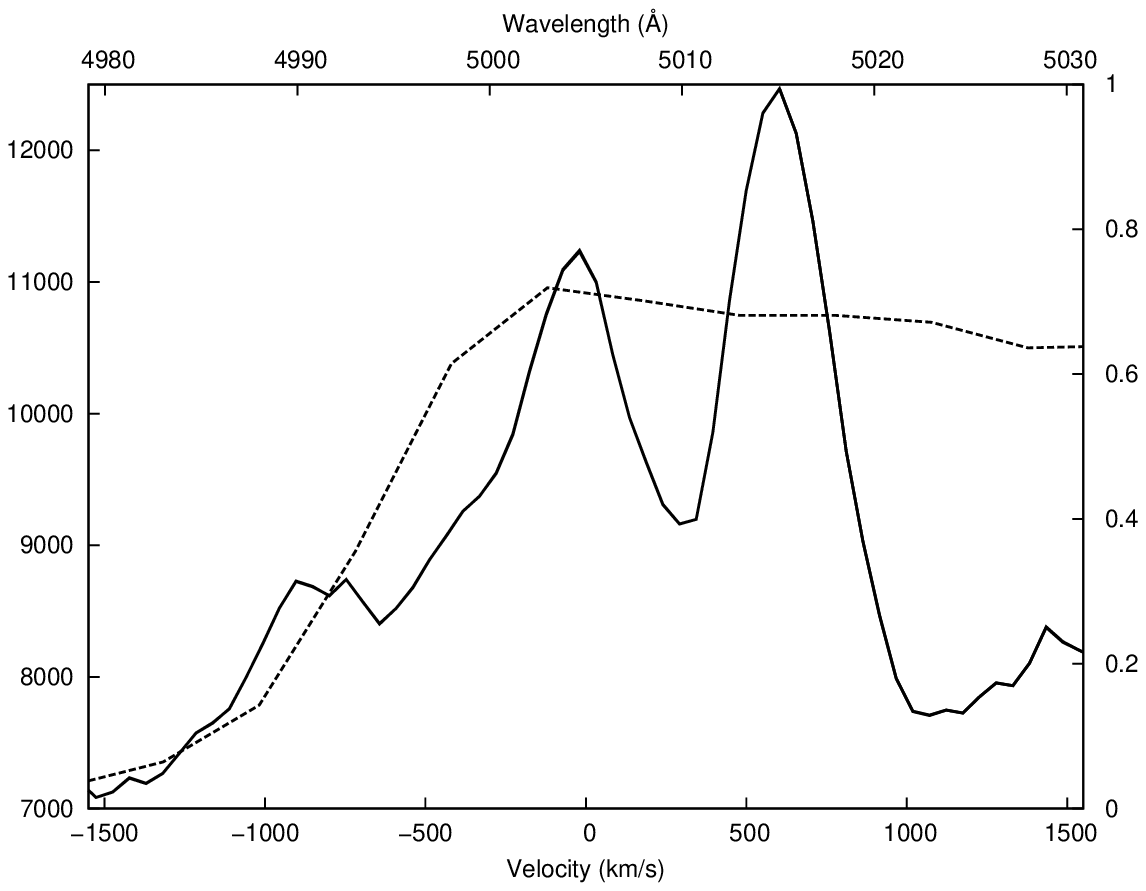}
\caption{{\it Top left} -- observed {\it HST} image on day 449 after outburst \citep[from][]{RBD09}. {\it Top right} -- synthetic image for the model where the outer component was allowed to expand linearly, while the inner component was kept the same physical size.  {\it Bottom} -- observed spectrum on day 415 after outburst (solid line) along with the {\it HST} F502N filter profile (dashed line). This demonstrates that the observed image, on the left, is dominated by a combination of the both N~{\sc ii} and He~{\sc i} $+$ Fe~{\sc ii}.}
\label{fig2}
\end{center}
\end{figure}

We identified the lines in the region around 5007 \AA, which at day 155 after outburst was dominated by the [O {\sc iii}] 5007\AA\ line, with a few lines of Fe II and He I 5018\AA\ arising, most likely, from the pre-existing red-giant wind. However, by the time of the spectroscopic observations, on day 415 after outburst, the lines around 5018 \AA\ are much broader and there is no evidence for [O {\sc iii}] 5007\AA\ confirmed by the absence of [O {\sc iii}] 4959\AA. This region is, by day 415, dominated by emission lines of N~{\sc ii} and He~{\sc i} + Fe~{\sc ii} (Figure~\ref{fig2}). We centre the velocity on N~{\sc ii} 5005\AA. Also shown in Figure~\ref{fig2} is the {\it HST} WFPC2 F502N image and filter profile. This aids in understanding what the main contributors to the observed emission on the {\it HST} image on day 449 after outburst are (Figure~\ref{fig2}).

\begin{figure}[t]
\begin{center}
\includegraphics[width=\textwidth]{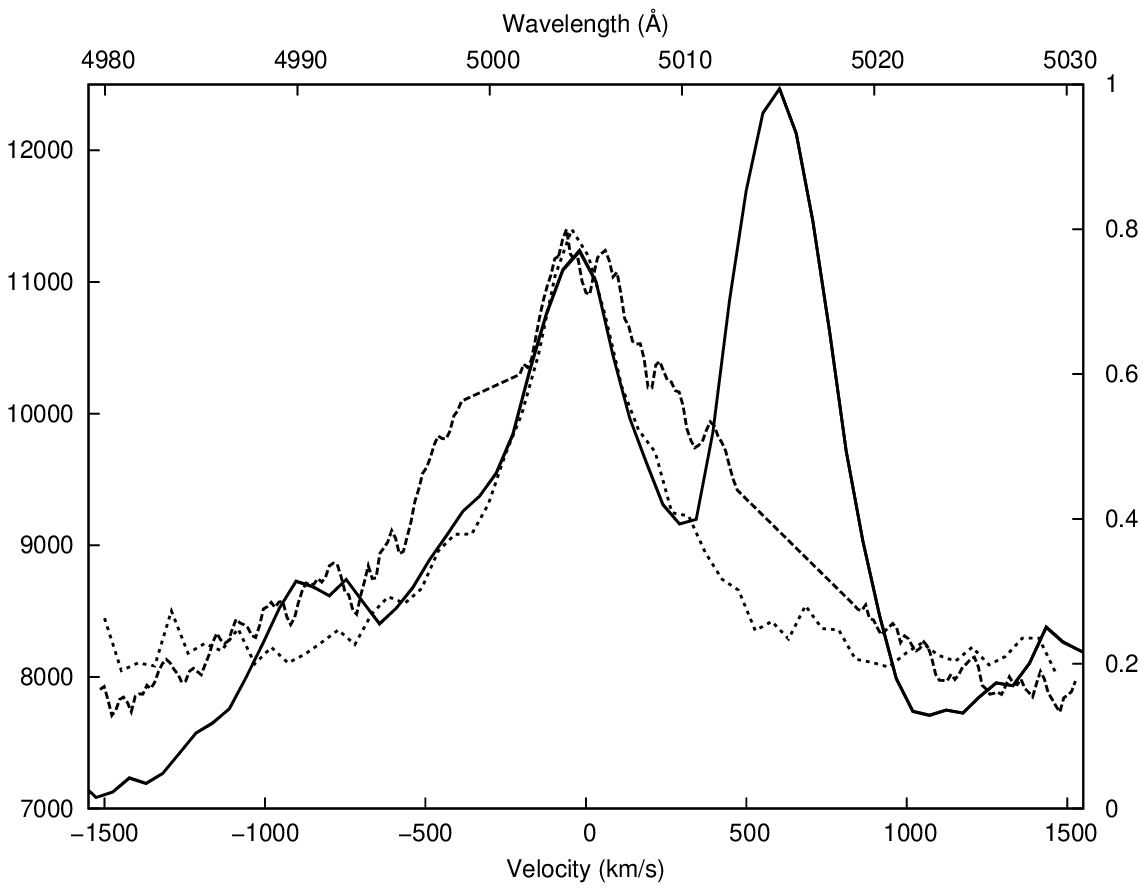}
\caption{The observed line profile at day 415 after outburst (solid line) along with model line profiles derived from the different morphologies; the linear expansion of all components (dashed line) and the linear expansion of the outer component while the inner component was kept the same size (dotted line). Note that detailed structure in the model line profiles is an artifact due to the finite resolution of the modelling grid.}
\label{fig3}
\end{center}
\end{figure}

We applied the morpho-kinematical code {\sc shape}\footnote{Available at: http://bufadora.astrosen.unam.mx/shape/} \citep{SKW11} using the morphologies from the day 449 results of \cite{RBD09}. Two models considered were: i) a linear expansion of both the inner and outer components, keeping inclination and axial ratios the same, which would produce the same line profile as for day 155; and ii) a model where the outer component was allowed to expand linearly, while the inner component was kept the same physical size. The results are shown in Figure~\ref{fig3}. What is clearly evident is that a model with a linear expansion of both components does not fit the observed line profile. To replicate the line profile, with model (ii), the innermost component was required to be much denser than the outer most material.

A comparison image derived from the model is shown in Figure~\ref{fig2}. The low resolution of the WFPC2 image prohibited a firm conclusion to be drawn at the time of writing the \citet{RBD09} paper. However, with the recently discovered WHT spectrum with which we can make a direct comparison, we can get closer to understanding the evolution of RS Oph's nebular remnant.

\section{Discussion and Conclusions}
The change of emission line with which the models of the structure of RS Oph were originally compared appears to make no difference in the overall derived morphology of the system. The absence of forbidden lines in the late-time spectra suggests emission may now be arising predominantly from dense clumps as the general trend is obviously for the density to drop with time. The morphology was well modelled with an outer dumbbell and an inner hour glass structure as on day 155. The inner structure unsurprisingly has a velocity less than the outer component and is much denser. This was required to replicate the observed low velocities seen in the lines. However, the innermost component appears to suffer significant deceleration compared with the outer component. We are now working on identifying all the lines in the archival spectra. Using a wide time span of spectra we can also track the evolution of the outburst (Ribeiro et al., in preparation).

There is weak indication of the presence of faint Fe~{\sc ii} (42) lines, and it is not surprising to speculate that this argues for two separate components of emitting gas. One is lower ionization and the other is the source of the He~{\sc i}. Furthermore, there is evidence of P~Cyg profiles emerging in the He~{\sc i} lines which may arise from the red-giant wind.

It is interesting to note the origin of the inner component. The potential survival of this dense material is also in line with evidence for the presence of silicate dust which survives the luminosity impulse and shock wave from the eruption \citep{EWH07}. The survival of the silicate dust may have implications regarding the formation mechanism of the remnant. Recently, 3D smoothed particle hydrodynamics modelling by \citet{MP12} and \citet*{MBP13} yielded a pre-outburst circumstellar density distribution that is highly concentrated towards the orbital plane and producing naturally the bipolar post-outburst outflow.

\acknowledgements
This paper makes use of data obtained from the Isaac Newton Group Archive which is maintained as part of the CASU Astronomical Data Centre at the Institute of Astronomy, Cambridge. The South African SKA Project is acknowledged for funding the postdoctoral fellowship position of VARMR at the University of Cape Town.


\begin{thebibliography}{}
\bibitem[{{Anupama} \& {Miko{\l}ajewska}(1999)}]{AM99}
{Anupama}, G.~C., \& {Miko{\l}ajewska}, J. 1999, \aap, 344, 177

\bibitem[{{Bode}(2010)}]{B10}
{Bode}, M.~F. 2010, Astronomische Nachrichten, 331, 160

\bibitem[{{Dobrzycka} \& {Kenyon}(1994)}]{DK94}
{Dobrzycka}, D., \& {Kenyon}, S.~J. 1994, \aj, 108, 2259

\bibitem[{{Evans} et~al.(2007){Evans}, {Woodward}, {Helton}, {van Loon},
  {Barry}, {Bode}, {Davis}, {Drake}, {Eyres}, {Geballe}, {Gehrz}, {Kerr},
  {Krautter}, {Lynch}, {Ness}, {O'Brien}, {Osborne}, {Page}, {Rudy}, {Russell},
  {Schwarz}, {Starrfield}, \& {Tyne}}]{EWH07}
{Evans}, A., {Woodward}, C.~E., {Helton}, L.~A., et al.
  2007, \apjl, 671, L157

\bibitem[{{Hounsell} et~al.(2010){Hounsell}, {Bode}, {Hick}, {Buffington},
  {Jackson}, {Clover}, {Shafter}, {Darnley}, {Mawson}, {Steele}, {Evans},
  {Eyres}, \& {O'Brien}}]{HBH10}
{Hounsell}, R., {Bode}, M.~F., {Hick}, P.~P., et al. 2010,
  \apj, 724, 480

\bibitem[{{Mohamed} et~al.(2013){Mohamed}, {Booth}, \& {Podsiadlowski}}]{MBP13}
{Mohamed}, S., {Booth}, R., \& {Podsiadlowski}, P. 2013, in Binary Paths to Type Ia Supernovae Explosions, edited by R. di Stefano, M. Orio, \& M. Moe (Cambridge: Cambridge Univ. Press), vol. 281 of the IAU Symposium, 195

\bibitem[{{Mohamed} \& {Podsiadlowski}(2012)}]{MP12}
{Mohamed}, S., \& {Podsiadlowski}, P. 2012, Baltic Astronomy, 21, 88

\bibitem[{{Narumi} et~al.(2006){Narumi}, {Hirosawa}, {Kanai}, {Renz},
  {Pereira}, {Nakano}, {Nakamura}, \& {Pojmanski}}]{NHK06}
{Narumi}, H., {Hirosawa}, K., {Kanai}, K., {Renz}, W., {Pereira}, A., {Nakano},
  S., {Nakamura}, Y., \& {Pojmanski}, G. 2006, IAU Circ., 8671

\bibitem[{{O'Brien} et~al.(2008)}]{OBB08}
{O'Brien}, T.~J., {Beswick}, R.~J., {Bode}, M.~F., et al. 2008, in
  RS Ophiuchi (2006) and the Recurrent Nova Phenomenon, edited by
  A. Evans, M. F. Bode, T. J. O'Brien, and M. J. Darnley (San
  Francisco: ASP), vol. 401 of the ASP Conf. Ser., 239

\bibitem[{{O'Brien} et~al.(2006){O'Brien}, {Bode}, {Porcas}, {Muxlow}, {Eyres},
  {Beswick}, {Garrington}, {Davis}, \& {Evans}}]{OBP06}
{O'Brien}, T.~J., {Bode}, M.~F., {Porcas}, R.~W., et al. 2006, Nature, 442, 279

\bibitem[{{Ribeiro}(2011)}]{R11}
{Ribeiro}, V.~A.~R.~M. 2011, Ph.D. thesis, Liverpool John Moores University

\bibitem[{{Ribeiro} et~al.(2009)}]{RBD09}
{Ribeiro}, V.~A.~R.~M., {Bode}, M.~F., {Darnley}, M.~J., et al. 2009, \apj, 703, 1955

\bibitem[{{Sokoloski} et~al.(2008){Sokoloski}, {Rupen}, \&
  {Mioduszewski}}]{SRM08}
{Sokoloski}, J.~L., {Rupen}, M.~P., \& {Mioduszewski}, A.~J. 2008, \apjl, 685,
  L137

\bibitem[{Steffen et~al.(2011)Steffen, Koning, Wenger, Morisset, \&
  Magnor}]{SKW11}
Steffen, W., Koning, N., Wenger, S., Morisset, C., \& Magnor, M. 2011, IEEE
  Transactions on Visualization and Computer Graphics, 17, 454, arXiv:1003.2012

\bibitem[{{Taylor} et~al.(1989){Taylor}, {Davis}, {Porcas}, \& {Bode}}]{TDP89}
{Taylor}, A.~R., {Davis}, R.~J., {Porcas}, R.~W., \& {Bode}, M.~F. 1989,
  \mnras, 237, 81

\bibitem[{{Yaron} et~al.(2005){Yaron}, {Prialnik}, {Shara}, \&
  {Kovetz}}]{YPS05}
{Yaron}, O., {Prialnik}, D., {Shara}, M.~M., \& {Kovetz}, A. 2005, \apj, 623,
  398

\end{thebibliography}
\end{document}